\documentclass[aps,prb,twocolumn,superscriptaddress]{revtex4-1}
\usepackage{graphicx}
\usepackage{epstopdf}
\usepackage{amsmath}
\usepackage{amsfonts}
\usepackage[usenames,dvipsnames]{color}
\usepackage{bm}
\usepackage{amsmath}

\usepackage{changes}
\usepackage[colorlinks=true,linkcolor=blue,urlcolor=blue,citecolor=blue]{hyperref}
\usepackage{lineno, blindtext}

\definecolor{pink}{rgb}{1,1,0} 
\definecolor{red}{rgb}{1,0,0}
\definecolor{yellow}{rgb}{1,1,0}
\definecolor{orange}{rgb}{1,0.5,0}
\definecolor{green}{rgb}{0,1,0}
\definecolor{blue}{rgb}{0,0,1}
\definecolor{white}{rgb}{1,1,1}
\definecolor{purple}{rgb}{0.5,0,0.5}

\begin{document}
\title{Topological edge states in dipolar zig-zag stripes}
\author{{ Paula Mellado$^{1,2}$}\\
{\small \em $^1 $School of Engineering and Sciences, 
	Universidad Adolfo Ib{\'a}{\~n}ez,
	Santiago, Chile\\
	\small \em $^2 $CIIBEC, Santiago, Chile.}}

\begin{abstract}
We study the magnon spectrum of stacked zig-zag chains of point magnetic dipoles with an easy axis. The anisotropy due to the dipolar interactions and the two-point basis of the zig-zag chain unit cell combine to give rise to topologically non-trivial magnon bands in 2d zig-zag lattices. Adjusting the distance between the two sublattice sites in the unit cell causes a band touching, which triggers the exchange of the Chern numbers of volume bands switching the sign of the thermal conductivity and the sense of motion of edges modes in zig-zag stripes. We show that these topological features survive when the range of the dipolar interactions is truncated up to the second nearest neighbors.
\end{abstract}

\maketitle

\section{Introduction}
\label{sec:intro}

In topological magnonics \cite{wang2018topological,zhang2013topological,mook2014edge,chisnell2015topological}, the hope of transmitting information unidirectionally, reliably, at specific frequencies and at a reduced dissipation rate seems achievable. The main players in charge of such a task are the spin waves, collective magnetic information which, due to the bulk-edge correspondence \cite{girvin2019modern}, are able of chiral propagation at the edges of those samples regardless of the specific device geometry. Nontrivial topology interesting by its own is usually accompanied by the emergence of other exotic phenomena \cite{kato2004observation,nagaosa2010anomalous,katsura2010theory} like in the insulating collinear ferromagnet $\rm{Lu_2V_2O_7}$ for instance,  where the spin excitations give rise to the anomalous thermal Hall effect \cite{nagaosa2010anomalous}. In this pyrochlore,  the propagation of the spin waves is influenced by the antisymmetric Dzyaloshinskii-Moriya (DM) spin-orbit interaction which plays the role of the vector potential. This is also the case in the compound $\rm{YMn_6Sn_6}$, a metallic system consisting of ferromagnetic kagome planes \cite{li2021dirac}, where the subsequent magnon band gap opening at the symmetry-protected  points is ascribed to DM interactions. Antiferromagnetic materials with honeycomb lattices add to the list of unique crystal structures which realize topological phases \cite{lee2018magnonic,owerre2017topological,PhysRevB.104.L060401}.  

In two dimensions (2D), magnonic crystals can host topological magnons when the system under consideration breaks time and space inversion symmetries \cite{shindou2013topological}. Evidence shows that this is possible when the system realizes spin-orbit physics on superlattices. In these structures, the common elements usually consist of units cells with the triangular motif and interactions that include all or some of the following elements: DM, anisotropy, exchange interactions, dipolar interactions, and external magnetic fields \cite{mook2014edge,chisnell2015topological,li2021magnonic,nikolic2020quantum}. Control over the band gaps and topological features in these structures include variations in magnetic fields, DM, temperature, and exchange interactions \cite{kim2016realization,pirmoradian2018topological,rau2016spin}. 

In systems where dipolar interactions play a dominant role, several proposals of chiral spin-wave modes in dipolar magnetic films have been put forward \cite{liu2020dipolar,pirmoradian2018topological, shindou2013topological,shindou2013chiral}, when subjected to an external magnetic field. Topological magnons also arise in magnonic crystals with the dipolar coupling truncated after a few neighbor magnets and in the case of antiferromagnetic films in the long-wave limit \cite{liu2020dipolar,pirmoradian2018topological}. In these systems, control over the dipolar magnonic bands is achieved through the application of magnetic fields.  

In this paper we study the magnon spectrum of two dimensional systems built from infinite zig-zag chains of point magnetic dipoles with an easy axis anisotropy. The two point basis of the zig-zag unit cell combine with the anisotropy of dipolar interactions between point dipoles to  give rise to topologically non trivial magnon bands in two dimensional zig-zag lattices. Tuning the distance $h$ between the sublattice sites  in the unit cell of the zig-zag chain causes a band touching, which triggers the exchange of the Chern numbers of the volume bands and switches the sign of the thermal conductivity in the 2d zig-zag lattices. Stripes made out of a finite number of zig-zag chains stacked along $\hat{y}$ exhibit a rich magnon spectrum where frequency gaps, thermal conductivity and the sense of motion of edges modes are easily manipulable by $h$. Finally, we find that thought the localization of the edges states deteriorates, the topological features of the systems survive when the range of the dipolar interactions is truncated up to the second nearest neighbors.  

The paper is organized as follows. In section ~\ref{sec:model} we present the model and the equilibrium magnetic configurations of the system. In this section we also demonstrate the mapping between dipolar interactions in the system and the symmetric and antisymmetric long ranged couplings. In section ~\ref{sec:waves} we show results for the magnon spectrum of the systems using the Landau-Lifshitz equations. Section ~\ref{sec:chern} is dedicated to the study of topological aspects of bulk bands and edge magnon modes in terms of $h$. In section ~\ref{sec:hall} we consider the magnon Hall effect in stripes while section ~\ref{sec:effective} is devoted to the study of an effective model for magnon spectrum in 1d systems. We summarize our results in Section ~\ref{sec:conclusion}. 
 \begin{figure}
  \includegraphics[width=\columnwidth]{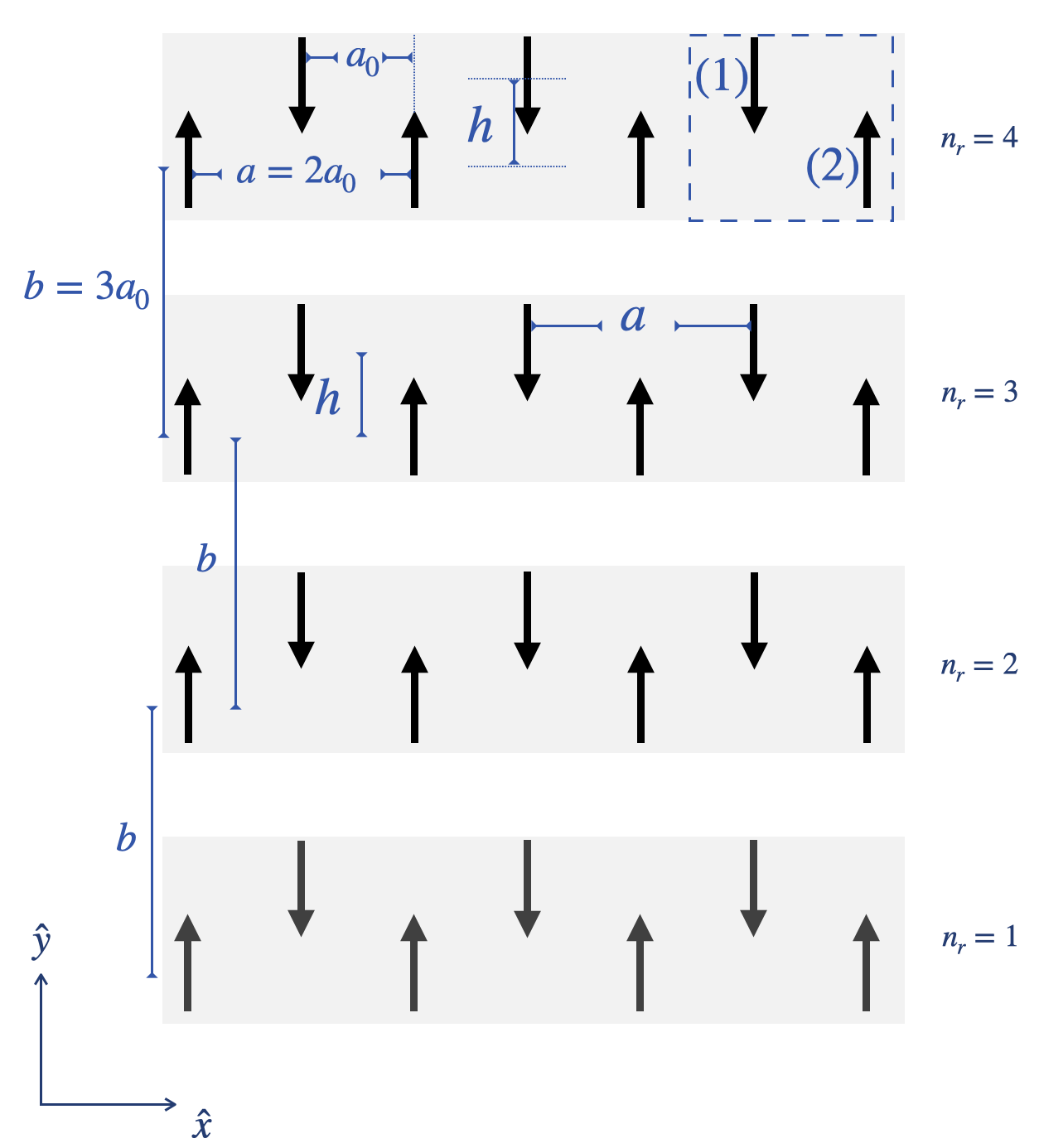}
\caption{Point dipoles are located in zig-zag chains: 1d lattices with a two point basis.  The dotted square highlights its unit cell that contains sublattices $(1)$ and $(2)$. $h$ is the tunable distance along $\hat{y}$ between sublattices (1) and (2) in a unit cell. Stacked zig-zag chains form stripes and 2d lattices with fixed lattice constants $a=2a_0$ and $b=3a_0$, along $\hat{x}$ and $\hat{y}$ respectively. $n_r$ indexes the rows along $\hat{y}$ in a stripe.}
\label{f1}
\end{figure}
\section{Model}
\label{sec:model}
The system energy consists of full ranged dipolar interactions and uniaxial anisotropy. In units of Joule $\rm[J]$ it reads 
\begin{eqnarray}
\mathcal{U}&=&\mathcal{U}_d+\mathcal{U}_K=\frac{\mathcal{\gamma}}{2} \sum_{i\neq k=1}^{N} \frac{\hat {\bm  m}_i \cdot\hat {\bm m}_k - 
3 (\hat {\bm m}_i \cdot \hat {\bm{e}}_{ik} )(\hat {\bm m}_k\cdot \hat {\bm{e}}_{ik} )}{|{\bm r}_i -{\bm r}_k |^3}+ \nonumber \\&&  - \,
\frac{{\mathcal K}}{2} \sum_{i=1}^N(\hat {\bm m}_k\cdot \hat {\bm{y}})^2,
\label{U}
\end{eqnarray}
where $N$ is the total number of sites and ${\bm r}_i$ denotes the position of a point dipole in a two dimensional system in the $x-y$ plane. $\hat {\bm{e}}_{ik}= ({\bm r}_i -{\bm r}_k) /|{\bm r}_i -{\bm r}_k|$, and $\mathcal{\gamma}=\frac{\mu_0 m_0^2}{4\pi a_0^3}$ has units of $[\rm Nm]$ and contains the physical parameters involved in the dipolar energy such as $\mu_0$ the magnetic permeability, $a_0$, the distance among nearest neighbor dipoles along the $\hat{x}$ direction, and ${m_0}$, the intensity of the magnetic moments with units $\rm[m^2 A]$. $\mathcal{K}$ is the easy axis anisotropy along the $\hat{y}$ axis and has units of $[\rm Nm]$. Magnets can rotate in an easy $x-y$ plane described by an azimuthal angle $\theta_i$ which is chosen with respect to the $\hat{y}$ axis. The magnetic moments of point dipoles have unit vector
$ \hat{{\bm m}}_i = (\cos\theta_i ,\sin\theta_i ,0)$. They are located at the vertices of infinite zig-zag chains that extend along $\hat{x}$. These are one dimensional (1d) lattices that extend along the $\hat{\bf x}$ direction, have lattice constant $a=2a_0$ and have a two point basis as shown in Fig.~\ref{f1}. The two sublattices are denoted by (1) and (2) while the unit cell is highlighted by the dotted square in Fig.~\ref{f1}. In the unit cell, $h$, the distance between dipoles (1) and (2) along $\hat{y}$ is a tunable parameter,  while $a_0$, the distance along $\hat{x}$ is kept fixed.  A finite number of stacked zig-zag chains form stripes and an infinite number of stacked zig-zag chains form zig-zag lattices with fixed lattice constants $a=2a_0$ and $b=3a_0$ along $\hat{x}$ and $\hat{y}$ axes respectively. 

Magnetic stable configurations are determined by the competition between the full dipolar interactions and the easy axis anisotropy that enforces dipoles to point along $\hat{y}$, as shown  respectively by the first and second terms in Eq.~(\ref{U}).  For small anisotropy, $\mathcal{K}<{\mathcal K}_c^{h}$, minimization of Eq.~(\ref{U}) favors a collinear magnetic order along $\hat{x}$. For $\mathcal{K}>{\mathcal K}_c^{h}$ parallel magnetic states along $\hat{y}$ minimize the full energy of 1d and 2d systems. From Eq.~(\ref{U}) the critical anisotropy that sets the boundary between collinear and parallel configurations is found to be ${\mathcal K}_c^{h}\sim2\gamma\left(\frac{3 \zeta(3)}{4}-\frac{1}{2} \psi^{(2)}\left(\sqrt{h^2+1}\right)+\frac{1}{2}\right)$, where $\psi$ is the PolyGamma function and $\zeta$ the Riemann zeta function (details can be found in the supplementary file \cite{supp}). 

In this paper we focus in the high anisotropy regime, $\rm {\mathcal K}>{\mathcal K}_c^{h}$. In this regime, for $h\lesssim a_0$ the antiferromagnetic parallel configuration of Fig.~\ref{f1} is favored in 1d and 2d systems. At intermediate amplitudes $1.5a_0\lesssim h\lesssim 2a_0$ dimers consisting of nearest neighbour ferromagnetic parallel dipoles arranged in an antiferromagnetic fashion `down-down-up-up' (and time reversal) along $\hat{y}$ are energetically favored \cite{supp}. 

Two symmetry notes call for attention: (1) when the distance $h=1a_0$,  single zig-zag chains becomes symmetric respect to an axis oriented in $\pi/4$ respect to the $\hat{x}$ axis (2) in stripes, at $h=1.5a_0$ the distance between two nearest  dipoles in chain $n_r$ is equal to the distance with its nearest neighbors in the closest zig-zag chain $n_{r\pm1}$, that is the stripe becomes symmetric respect to the $\hat{x}$ axis. Finally we note that in 2d systems, at  $h>1.5a_0$, the stronger dipolar couplings occurs between neighbor dipoles that belong to different zig-zag chains.  

Hereinafter the length scales are in units of $a_0$ and the energy scales in units of $\gamma/2$.

\subsection{Exact mapping of $\mathcal{U}_d$, to isotropic and anisotropic couplings}
\label{sec:mapping}
In Eq.\ref{U}, the dipolar coupling  $\mathcal{U}_d$ maps exactly into the hamiltonian $\mathcal{H}_d$,
\begin{widetext}
\begin{align}
\mathcal{H}_d&=&\frac{\mathcal{\gamma}}{2} \sum_{i\neq k=1}^{\frac{N}{2}}\sum_{s=1}^2 \left[J_1\left(-\hat {\bm  m}_i^{s} \cdot\hat {\bm m}_k^{s}+3\cos{(\theta_i^{s}+\theta_k^{s})}\right)+ \sum_{t=1}^2 \left[-J_2\hat {\bm  m}_i^{s} \cdot\hat {\bm m}_k^{t}+J_3\cos{(\theta_i^{s}+\theta_k^{t})} +\mathcal{\bm D}\cdot (\hat {\bm  m}_i^{s}\times\hat{\bm m}_k^{t})\right]\right],
\label{eq2}
\end{align}
\end{widetext}
where $s$ and $t$ index sublattices (1) and (2) (Fig.\ref{f1}), $J_1=\frac{1}{2|i-k|^{3}}$, $J_2=\frac{1}{2((i-k+\frac{1}{2})^{2}+h^{2})^{\frac{3}{2}}}$, $J_3=\frac{-3(h^{2}-(i-k+\frac{1}{2})^{2})}{2((i-k+\frac{1}{2})^{2}+h^{2})^{\frac{5}{2}}}$ are respectively symmetric long range couplings between dipoles in the same sublattice ($J_1$) and dipoles in different sublattices ($J_2$ and $J_3$). Finally $\mathcal{\bm D}=\frac{-3h(i-k+\frac{1}{2})}{((i-k+\frac{1}{2})^{2}+h^{2})^{\frac{5}{2}}}\hat{z}$ is an antisymmetric coupling between dipoles in distinct sublattices. The antisymmetric interaction is of the spin orbit type and is interpreted as a long range Dzyaloshinkii-Moriya coupling in the system. 
 \section{Magnon spectrum}
\label{sec:waves}
Next we examine the collective transverse excitations of the magnetization vector with respect to the parallel magnetic states.  Neglecting damping, the dynamics of the magnetization vector of a dipole $\jmath$, belonging to sublattice $s$, $\bm {\hat{m}}_j^{s}(t)$ is described by the  Landau-Lifshitz equation \cite{lakshmanan2011fascinating,osokin2018spin,galkin2005collective,bondarenko2010collective,verba2012collective,lisenkov2014spin,lisenkov2016theoretical}
\begin{eqnarray}
\frac{d{\bm {\hat{m}}}_j^{s}}{dt} =g({\bf B}_{\rm eff,j}^{s}\times {\bm {\hat{m}}}_j^{s})
\label{eq:LL}
\end{eqnarray}
 where $g$ is the modulus of the gyromagnetic ratio, and the effective magnetic field ${\bm B}_{\rm {eff,j}}^s=-\frac{\partial {\mathcal H}_{d}}{\partial \bm {\hat{m}}_{j}^s}$
consists of the dipolar field created by other magnetic dipoles,
\begin{eqnarray}
{\bm B}_{\rm eff,j}^{s}=-\mu_0\sum_{i,t}\bm{\mathcal{I}}_{ij}^{s,t}\cdot{\bm{\hat{m}}_i^{t}},
\label{eq:eff}
\end{eqnarray}

\begin{figure}
\includegraphics[width=\columnwidth]{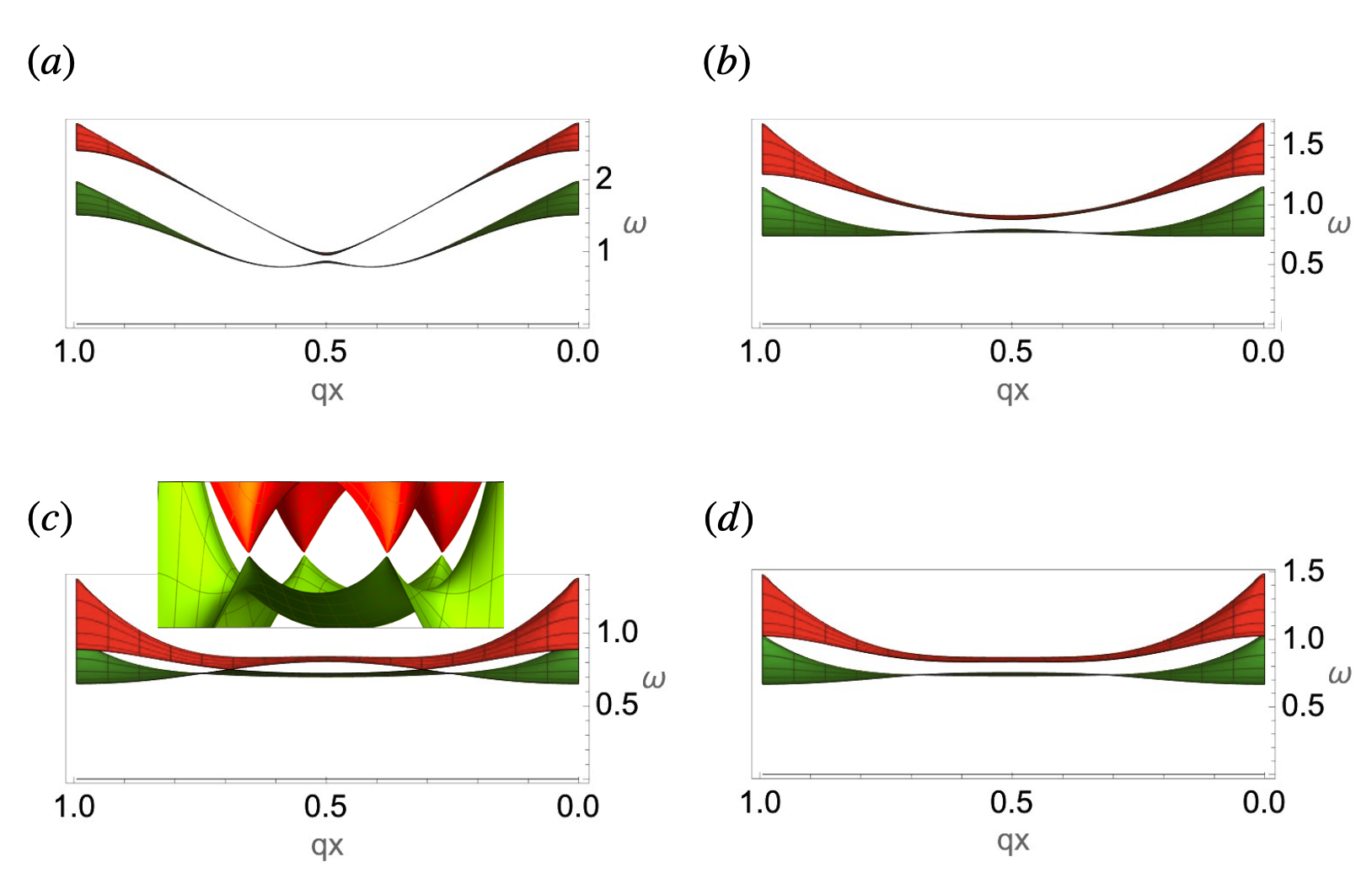}
\caption{Magnon spectrum of 2d lattices (side view at $q_y=0$) with  (a) $h=0.5$, (b) $h=1$, (c) $h=1.5$, and (d) $h=1.8$. In (c) two band touchings like Dirac points are located at $\left(\frac{G_x}{3},\frac{G_y}{2}\right)$ and $\left(\frac{2 G_x}{3},\frac{G_y}{2}\right)$ (and the two equivalent at $q_y=\frac{3G_y}{2}$) when $h=\frac{3}{2}$.  Frequency $\omega$ is in units of $g$ and $q_x$ is in units of $G_x$.}
\label{f2}
\end{figure}
\begin{figure}
\includegraphics[width=\columnwidth]{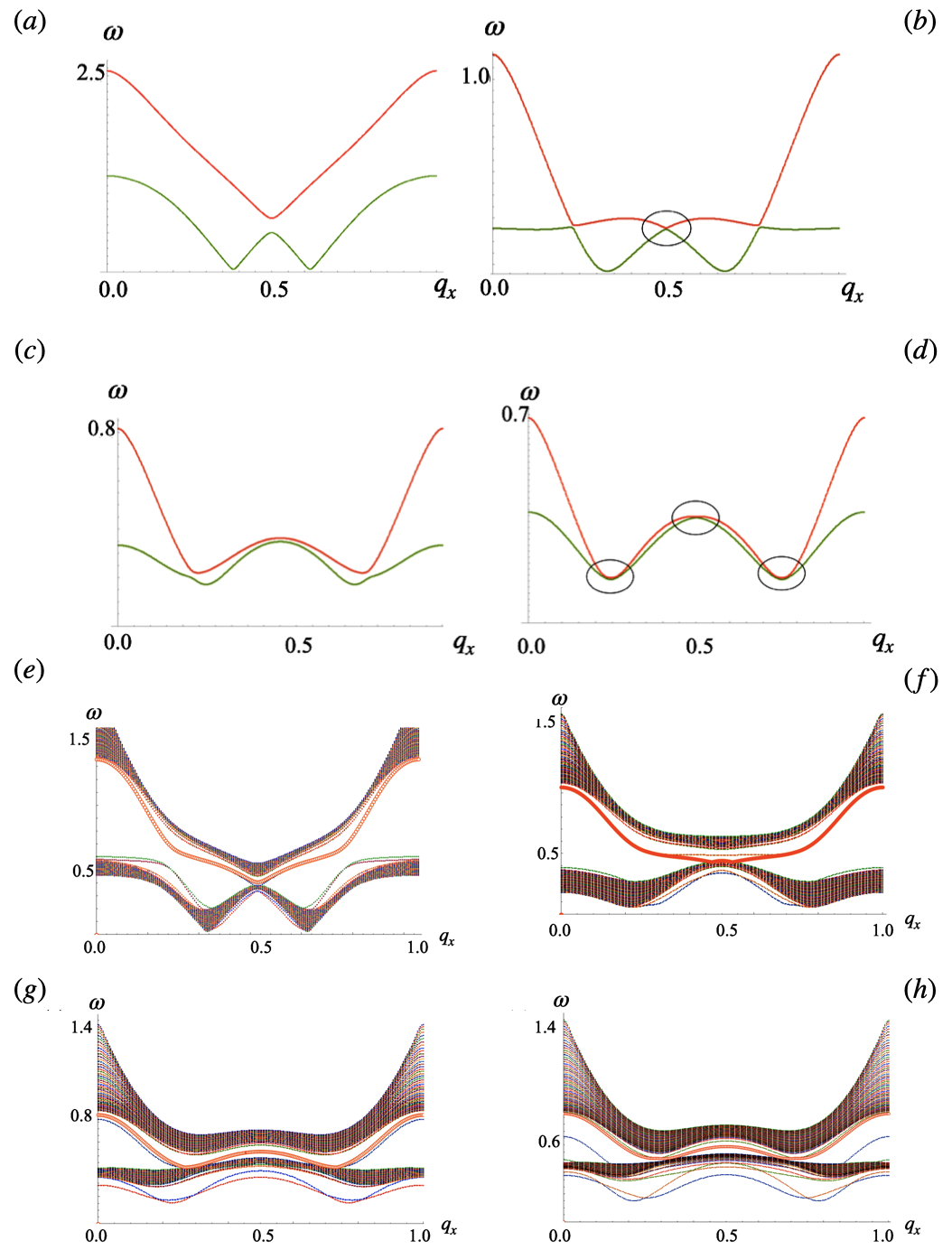}
\caption{(a-d) Magnon spectrum of zig-zag chains ($N_r=1$) at (a) $h=0.5$, (b) $h=1.0$, (c) $h=1.3$ and (d) $h=1.5$. (e-h) Magnon spectrum of stripes with $N_r=50$ at (e) $h=0.8$, (f) $h=1.0$, (g) $h=1.3$ and (h) $h=1.6$. Edge modes crossing the gap and joining the two bulk bands are highlighted. $\omega$ is in units of $g$ and $q_x$ is in units of $G_x$.}
\label{f3}
\end{figure}
where $\bm{\mathcal{I}}_{ij}^{s,t}$ is the interaction matrix containing the geometrical aspects of the dipolar coupling between all dipoles. Hereafter we drop the hat from unit vectors.
Magnetization of the $j$th dipole in the stationary ground state ${\bm\mu}_j^{s}=(0,0,\mu_{j}^{s})$ is a unit vector in the direction of the static ground state. It points along a local axis which we call  $\bm{\hat{z}}$ and satisfies the system of equations:
\begin{eqnarray}
B_j^s {\bm\mu}_j^s= -\mu_0\sum_{i,t}\bm{\mathcal {I}}^{s,t}_{z}\cdot{\bm\mu}_i^{t}
\label{eq:static}
\end{eqnarray}
where $B_j^s$ is the intrinsic scalar magnetic field acting on the $j$th dipole belonging to the $s$ sublattice and $\bm{\mathcal {I}}^{s,t}_{z}$ is the interaction matrix of the system in the stationary ground state. To find the dynamical equations describing small (linear) transverse
magnetization excitations respect to ${\bm\mu}_j^{s}$, we use the following ansatz for the dipole magnetization:
\begin{eqnarray}
\bm{m}_j^{s}(t)=({\bm\mu}_j^{s}+{ \tilde{\bm{m}}}_j^{s}(t))
\label{eq:linear}
\end{eqnarray}
where $\tilde{\bm {m}}_j^{s}(t)=(\tilde{m}_{j,x}^{s}(t),\tilde{m}_{j,y}^{s}(t))$ is the small dimensionless deviation of the magnetization vector of the $j$th dipole from the static equilibrium state. Conservation of the length of the magnetization vector in each magnet requires that ${\bm\mu}_j\cdot \tilde{{\bm m}}_j=0$. Equilibrium orientations of the uniform magnetization and internal fields depend only on the index $s$. At each sublattice, the linear spin-wave excitations have the form of plane waves, therefore we Fourier transform the magnetic excitation vector in time and space.
In terms of magnon creation $m_+=m_x+ i  m_y$ and annihilation $m_-=m_x- i m_y$ fields, the matrix form of the equations of motion becomes \cite{supp}
  \begin{eqnarray}
\omega \hat{\sigma_z}\left(\begin{array}{c}m_+ \\m_-\end{array}\right)=\left(\begin{array}{cc}\hat{A} & \hat{B} \\\hat{B} & \hat{A}\end{array}\right)\left(\begin{array}{c}m_+ \\m_-\end{array}\right)
\label{eq:magnon}
\end{eqnarray}
where the right hand side matrix constitutes the magnon hamiltonian. The Pauli matrix  $\hat{\sigma}_z$, takes $+ 1$  for the creation field or particle space and $- 1$ for the annihilation field or hole space.
The $2\times 2$ matrices $\hat{A}=\left(\begin{array}{cc}a_1 & a_2 \\a_2^* & a_1\end{array}\right)$ and $\hat{B}=\left(\begin{array}{cc}b_1 & b_2 \\b_2^* & b_1\end{array}\right)$ with $a_1=g \epsilon_0+\frac{g}{2}(I_{{\bm q},x}^{(11)}+I_{{\bm q},y}^{(11)})$, $a_2=g \epsilon_0+\frac{g}{2}(I_{{\bm q},x}^{(12)}+I_{{\bm q},y}^{(12)})$, $b_1=\frac{g}{2}(I_{{\bm q},x}^{(11)}-I_{{\bm {\bm q}},y}^{(11)})$, $b_2=\frac{g}{2}(I_{{\bm q},x}^{(12)}-I_{{\bm q},y}^{(12)})$ and $\epsilon_0$ is the energy of the stationary magnetic state (hereafter we set $\epsilon_0=0$). Expressions for $I_{{\bm q},x}^{11}$ and $I_{{\bm q},y}^{12}$ are found in \cite{supp}. Finally, the magnon hamiltonian becomes \cite{mellado2022intrinsic}:
\begin{equation}
\mathcal{H}_d=\hat{\sigma}_z\otimes(\hat{t}_0+\hat{t}_1)+\hat{\sigma}_x\otimes(\hat{t}_2+\hat{t}_3)
\label{eq:hmag}
\end{equation}
where $\hat{\sigma}_j$ is the jth Pauli matrix, $\otimes$ denotes Kronecker product, $\hat{t}_0=\left(\begin{array}{cc}a_1 & 0 \\0 & a_1\end{array}\right)$, $\hat{t}_1=\left(\begin{array}{cc}0 & a_2 \\a_2^* & 0\end{array}\right)$, $\hat{t}_2=\left(\begin{array}{cc}b_1 & 0 \\0 & b_1\end{array}\right)$ and $\hat{t}_3=\left(\begin{array}{cc}0 & b_2 \\b_2^* & 0\end{array}\right)$.
In Eq.~(\ref{eq:hmag}) the term multiplying $\hat{\sigma}_z$ is a mass term responsible for the gap in the magnon spectrum,  while the term multiplying $\hat{\sigma}_x$ is proportional to the group velocity of the spin waves or magnon speed.
Eigenfrequencies for collective spin wave modes in the particle space can be written as
\begin{equation}
\omega_{1,2}^2=\frac{g^2}{4}(a_1^2 \pm 2a_1a_2+a_
2^2+b_1^2 \pm 2b_1b_2+b_2^2)
\label{eq:freq1}
\end{equation}
Figs.~\ref{f2},~\ref{f3} show the effect of $h$ on the magnon spectrum in terms of $q_x$ the wavevector along $\hat{x}$ in the first Brillouin zone (1BZ), for 2d lattices, single zig-zag chains and stripes made out of $N_r=50$  zig-zag chains respectively. The spectrum is the result of the exact diagonalization of Eq.~(\ref{eq:hmag}) in each case.
A first apparent effect of reducing $h$ is the amplification of the range of frequencies that allow for magnon excitations at the $\Gamma$ point  in all cases.   In addition, the heigh $h$ has a strong effect on the group velocity of the spin waves: larger values of $h$ yield flatter bands and therefore lower magnon speeds around the middle of the spectrum: compare, for instance, frequency slopes of systems with $h=0.5$ and $h=1.8$ in Figs.~\ref{f2}(a) and (d) respectively.
Notable features are the band touchings in chains and lattices. In single chains, the first touching occurs in the middle of the spectrum $q_0=\frac{G_x}{2}$ at the symmetric point $h=1\equiv h_1$, Fig.~\ref{f3}(b). At $h>h_1$ the gap opens again until $h=1.5\equiv h_2$ where the touching at $q_0$ reappears and two additional band touchings arise at $q_1\sim\frac{G_x}{4}$ and $q_2\sim 3\frac{G_x}{4}$, Fig.~\ref{f3}(d). For $h>h_2$ the gap at $q_1$ and $q_2$ remains closed while that at $\frac{G_x}{2}$ reopens. One can estimate the critical amplitude for which the magnonic gap closes by equating the eigenfrequencies $\omega_1$ and $\omega_2$ in Eq.~(\ref{eq:freq1}). At $q_x=\frac{G_x}{2}$ this yields the equality
\begin{equation}
h^2_{q_x=\frac{G_x}{2}}=\frac{1}{2}\frac{\Phi \left(\imath,3,\sqrt{1+h^2}\right)}{\Phi \left(\imath,5,\sqrt{1+h^2}\right)}
\end{equation}
which is satisfied at $h\sim h_1$.

In the case of 2d lattices the band touching occurs at $h_2$ at two points of the 1BZ, $\left(\frac{G_x}{3},\frac{G_z}{2}\right)$, $\left(2\frac{G_x}{3},\frac{G_z}{2}\right)$ as shown in Fig.~\ref{f2}(c). Here the inset shows a close up of the band spectrum about the band touching points (and the equivalent $\left(\frac{G_x}{3},3\frac{G_z}{2}\right)$ and $\left(2\frac{G_x}{3},3\frac{G_z}{2}\right)$ ) which resembles Dirac cones. At all other $h$ the spectra of the lattices are gapped. We remind the reader that $h_2$ is a special symmetric point where interchain and intrachain nearest neighbor distances become equal.

In stripes, Fig.~\ref{f3}(e-h) shows magnon bands that cross the gap between the two bulk bands (highlighted in orange). Isolated bands located above and below the bulk bands arise as well for large $h$, as for instance the ones shown below the lowest frequency band in Fig.~\ref{f3}(h). In the following sections we examine the nature of the in gap eigenmodes.

 \section{Topological Bands and Edge Modes.}
 \label{sec:chern}
The in-gap magnon modes, apparent in the magnon spectrum of Fig.~\ref{f3}  resemble edge modes. Aimed to unveil their topological nature,  we examine the topological Chern number of the volume bands associated to them. A non-zero Chern integer, $c_n$, for spin-wave volume bands results in the emergence of chiral spin-wave edge modes. These topological edge bands have chiral dispersion that favors the unidirectional propagation of magnetic degrees of freedom for a frequency in the gap. In addition, they are robust to intrinsic and externally induced disorder \cite{girvin2019modern}.  
 It is expected that finite Chern integers, $c_n$, result from strong spin-orbit coupled interactions \cite{shindou2013chiral,peter2015topological}. With an inner product between magnetization and position vectors, the magnetic dipolar coupling locks the relative rotational angle between the spin space and orbital space, similar to what the relativistic spin-orbit interaction does in electron systems \cite{shindou2013chiral}. Indeed, this is the case in our zig-zag system as shown in  Eq.\ref{eq2} by the direct mapping between dipolar interactions  and a long ranged DM coupling between dipoles along $\hat{z}$. As a result of the spin-orbit locking, the complex-valued character in the spin space is transferred into wave functions in the orbital space. Consequently,  a topological magnonic spectrum is expected.
\subsection{Gauge connection and the associated Berry curvature in the lattice.}
 \label{sec:con}
The sense of motion and the number of chiral modes in a system is determined from the magnitude and sign of the topological number $c_n$ for volume mode bands below the bandgap. $c_n$, can be changed only by closing the gap \cite{girvin2019modern,shindou2013chiral}. 
\begin{table}[bt]
\begin{tabular}{|c|c|c|c|c|c|c|c|c|c|c|}
\hline
$h$&0.2&0.5&0.7&1&1.2&1.4&1.5&1.6&1.8&2.0\\
\hline
$c_1$&1&1&1&1&1&1&1&-1&-1&-1\\ 
\hline
$c_2$&-1&-1&-1&-1&-1&-1&-1&1&1&1\\ 
\hline
\end{tabular}
\caption{Chern numbers, of the magnon volume bands in 2d lattices made out of zig-zag chains at several values of the heigh $h$. Note the exchange of the Chern number between the two bands at the symmetric point $h_2=1.5$.}
\label{table:1}
\end{table}
\begin{figure*}
\includegraphics[width=\textwidth]{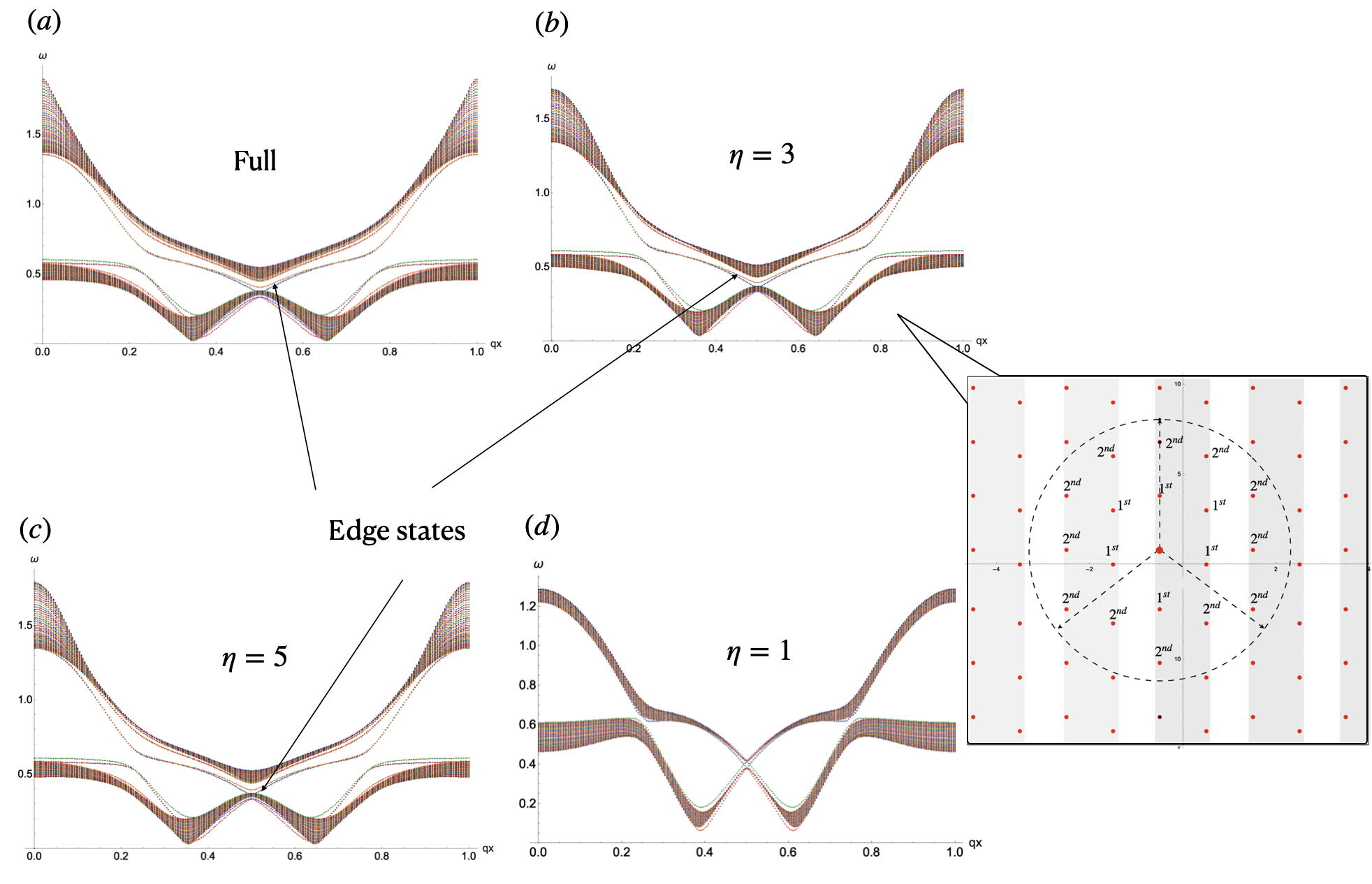}
\caption{Magnon spectrum of stripes with $N_r=50$ and heigh $h=0.8$. We examine the effect of the cutoff of the range of dipolar interactions in the system: a given dipole interact with all dipoles located inside a circle of radius $r=\eta a$, with $a$ the lattice constant. In the rightest illustration we show the case of $\eta=3$ which includes all the six first and twelve second nearest neighbor dipoles of the dipole highlighted by the central larger red dot. In (a-d) we compare respectively the magnon spectrum of the system with full ranged interactions (a) and $\eta=3$ (b), $\eta=5$ (c) and $\eta=1$ (d).}
\label{f4}
\end{figure*}
\begin{figure}
\includegraphics[width=\columnwidth]{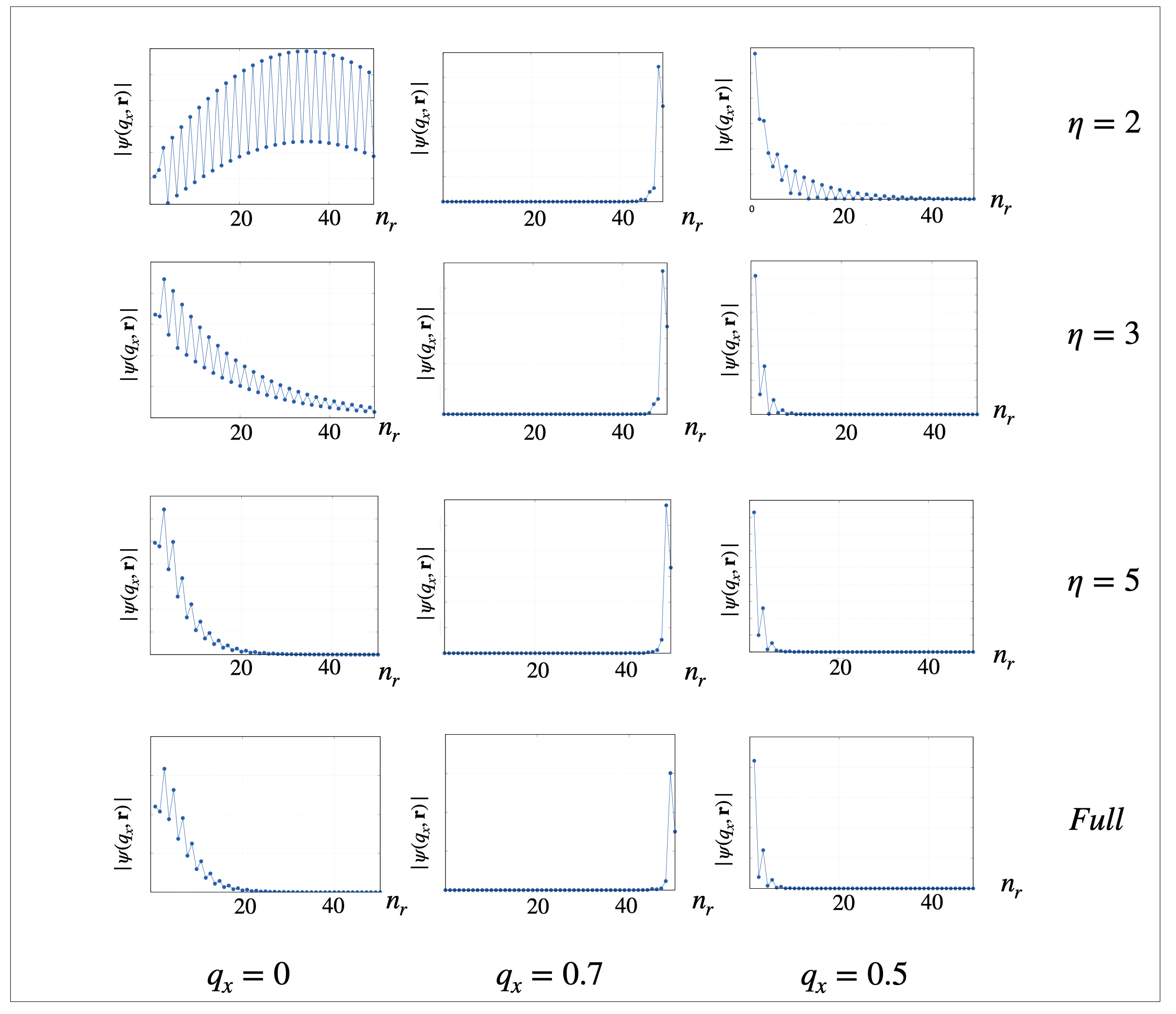}
\caption{Amplitude of the Bloch wave function of the in gap eigenmodes of the magnon spectrum of stripes with $N_r=50$ and heigh $h=0.8$ at the points of the 1BZ, $q_x=0$, $q_x=0.7$ and $q_x=0.5$. From the upper to the lower row we compare respectively the cases of the system with truncated dipolar interactions and $\eta=2$ (first row), $\eta=3$ (second), $\eta=5$ (third) and the full ranged interactions  case (fourth row).}
\label{f5}
\end{figure}
\begin{figure}
\includegraphics[width=\columnwidth]{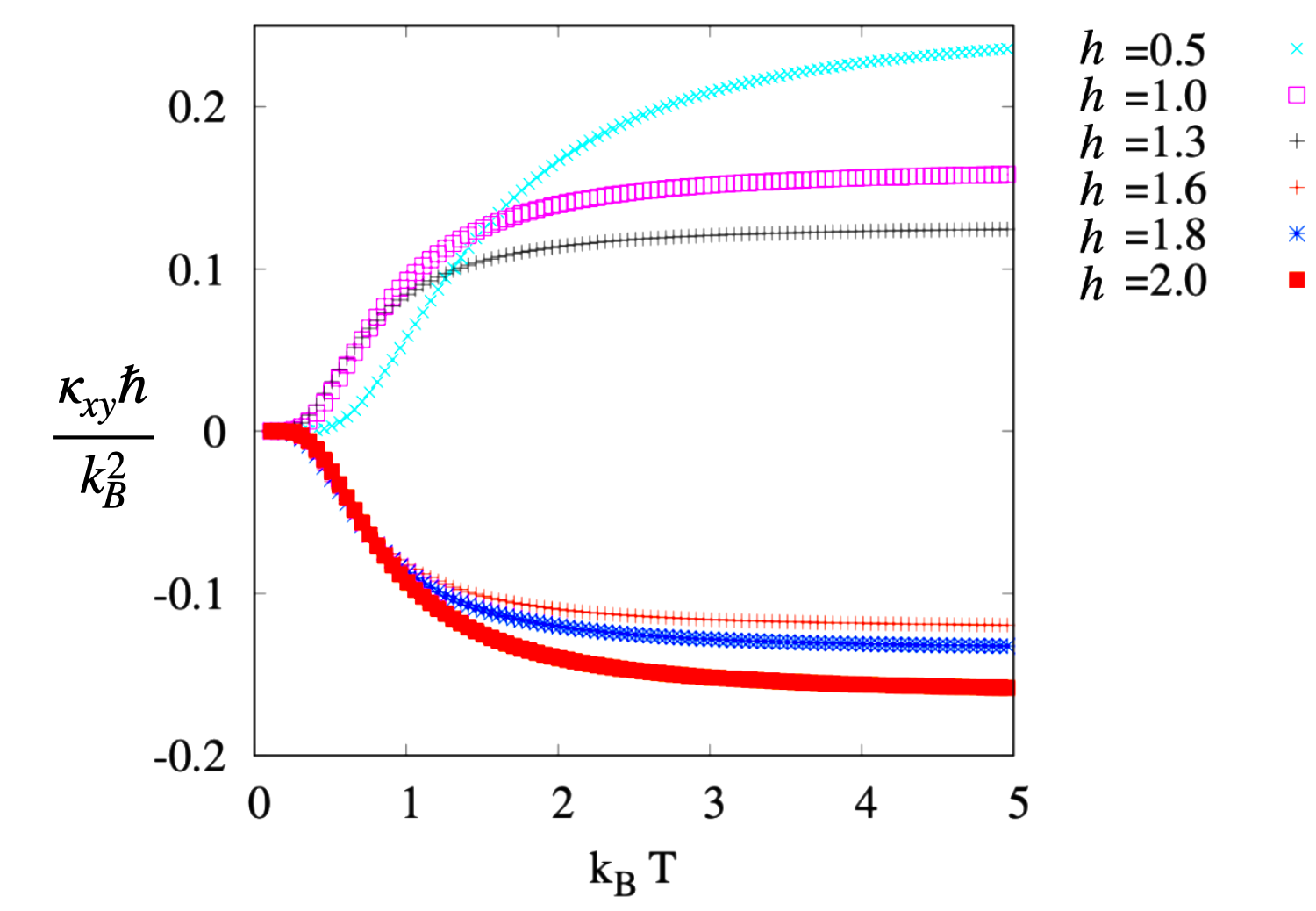}
\caption{Thermal conductivity as a function of $k_{B}T$ at different $h$. Note the change in the sign of $\kappa^{xy}$ for $h>1.5$.}
\label{f6}
\end{figure}
Here, frequency bands are computed on a discretized Brillouin zone. Thus, we follow the approach of reference \cite{fukui2005chern} to compute the Berry phase \cite{zak1989berry} and the Chern integer using wave functions given on such discrete points.
The Chern number assigned to the $n$th band is the integral of fictitious magnetic fields: that is, field strengths of the Berry connection. On the lattice, it is defined by  \cite{fukui2005chern}
  \begin{eqnarray}
 \tilde{c}_n\equiv\frac{1}{2\pi i}\sum_l \tilde{F}_{xy}(q_l)
 \label{eq:chern2}
\end{eqnarray}
The gauge connection (gauge field) $A_\mu(q)$ ($\mu=x,y$) and the associated field strength $F_{xy}(q)$ or Berry curvature are given by  
  \begin{eqnarray}
A_\mu&=&\langle n(q)|\partial_\mu| n(q) \rangle \\\nonumber
F_{xy}&=&\partial_xA_y(q)-\partial_yA_x(q)
 \label{eq:berry}
\end{eqnarray}
where $| n(q) \rangle$ is the normalized wave function of the $n$th (particle) Bloch band such that $H(q)| n(q) \rangle=\omega_n(q)| n(q) \rangle$.  
On the discrete Brillouin zone with lattice points $q_l$, $l=1,...n_xn_y$ the lattice field strength is given by 
  \begin{align}
\tilde{F}_{xy}(q_l)\equiv\ln U_x(q_l)U_y(q_l+\hat{x})U_x(q_l+\hat{y})^{-1}U_y(q_l)^{-1}\\\nonumber
-\pi<\frac{1}{i}\tilde{F}_{xy}(q_l)\leq \pi \nonumber
 \label{eq:discberry}
\end{align}
where $U_\mu(q_l)$ is a U(1) link variable from a $n$th band,
  \begin{eqnarray}
U_\mu(q_l)\equiv \frac{\langle n(q_l)| n(q_l+\hat{\mu}) \rangle}{|\langle n(q_l)| n(q_l+\hat{\mu}) \rangle|}
 \label{eq:link}
\end{eqnarray}
\subsection{Topological magnons at the edges of stripes}
\label{sec:edges}
Using the previous approach, we computed the Chern numbers for lattices (Fig.\ref{f2}) corresponding to the bulk bands of the stripes (Fig.\ref{f3}) with $N_r=50$ at several values of $h$. The results are presented in Table~\ref{table:1}. For $h\leq h_2$ the lowest frequency band has Chern number $c_1=1$ and the upper band has $c_2=-1$. This topological phase of the magnon dispersion is characterized in terms of $c_1$ and $c_2$ and denoted (1,-1). At $h_2$  the Chern numbers of the two bands are exchanged, and at $h>h_2$ the system realizes the topological phase denoted (-1,1).
Going back to Fig.~\ref{f2}(c), we note two band touchings at $h=h_2$, at points $(\frac{G_x}{3},\frac{G_y}{2})$ and $(\frac{2G_x}{3},\frac{G_y}{2})$ of the 1BZ. There, the bands form approximated gapless Dirac spectra (inset in Fig.~\ref{f2}(c)). 
A band touching point in the 3D parameter space $(q_x,q_y,h)$ plays the role of a dual magnetic monopole. The corresponding dual magnetic field is a rotation of the three component gauge field $\bm A_n=(A_{n,x},A_{n,y},A_{n,h})$, $\bm{B}_n=\bm\nabla\times \bm A_n$, where $\bm\nabla\equiv (\partial_{qx},\partial_{qy},\partial_h)$ and $n$ specifies either of the magnonic bands which form the band touching.  At the band touching point, the dual magnetic field for the respective bands has a dual magnetic charge, whose strength is quantized to be $2\pi$ times an integer \cite{girvin2019modern}. Because the Chern integer $c_n$ can be regarded as the total dual magnetic flux penetrating through the constant $h$ plane, the Gauss theorem implies that when $h$ goes across the $h=h_2$ plane, the Chern integer for the lowest magnonic band $c_1$ changes by unit per each touching point. Hence, due to the two band touchings $c_1|_{h>h_2}-c_1|_{h<h_2}=2$, which explains the exchange of Chern numbers between bands at $h_2$.  
According to the bulk-edge correspondence principle \cite{girvin2019modern} the number of in gap one-way edge states is determined by the winding number of a given band $n$, that is, the sum of all the Chern numbers of the band up to band $n$. Consequently, stripes in Fig.\ref{f3} should realize one topological edge mode at each edge. 

To further investigate these edge modes, we studied the localization of the Bloch wave function $\psi(q_x,\bm{r})$ associated to the edge state of the stripe with $h=0.8$ and $N_r=50$ (in gap mode highlighted in orange in Fig.\ref{f3}(e)) at three positions of the 1BZ, $q_x=0,0.7,0.5$. The three plots shown in the first row of Fig.\ref{f5}  (from bottom to top) show $|\psi(q_x,\bm{r})|$ in terms of the row number $n_r$ of the stripe. Row numbers grow from $n_r=1$ to  $n_r=50$.  At the three wavevectors  the amplitude of the Bloch function $|\psi(q_x,\bm{r})|$ is maximum at the edge of the stripe (which is at row $n_r=1$) and decreases in an exponential fashion as $\bm r$ progresses into the bulk of the stripe. Small oscillations can be seen at $q_x=0$ and $q_x=0.5$ which for $\eta\leq 5$ are associated to some degree of delocalization and hybridization of the edge mode with bulk states at $q_x=0$.
\subsection{Truncated dipolar interactions}
\label{sec:trun}
The topological nature of the edges states found at $h>0$ is due to the dipolar coupling between magnets which results in a long ranged DM kind of interaction in the system at hand. One natural question is whether or not the topological nature of the edge states survive when the range of the dipolar interactions is truncated.  Back into Eq.\ref{eq2} we note that such truncation will not only affect the range of DM interactions but also truncate the  symmetric exchange couplings in $\mathcal{H}_d$. We delay for a separated work the problem of finding the specific contribution of each coupling in Eq.\ref{eq2} to the topology of the system, and focus now in the qualitative behavior of the magnon spectrum under truncation of the dipolar interactions. For this purpose we consider the scenario where each dipole interact only with those magnets that are contained inside a circle of radius $r=\eta a$ around it: this is illustrated in the rightmost sketch of Fig.\ref{f4} for the case $\eta=3$. In this analysis, stripes with $N_r=50$ zig-zag rows and $h=0.8$ are considered, and the natural number $\eta$ has been variated from 1 in unity steps.  
The magnon spectrum for the cases of full interactions (Full), $\eta=3$, $\eta=5$ and $\eta=1$ is compared in Fig.\ref{f4}. We see a couple of magnon modes crossing the gap in each case. Of those, the one at higher frequency in orange  is considered here. For $\eta=1$, Fig.\ref{f4}(d) a band crossing  between the two in gap modes appears at $q_x=0.5$. In addition, band touchings between in gap modes and bulk modes are apparent at $q_x=0.5$ and $q_x=0$. However once $\eta\geq 3$, Fig.\ref{f4}(b) the band crossing and band touching at $q_x=0.5$ disappear. At $q_x=0$, for the full interactions shown in Fig.\ref{f4}(a) it is possible to distinguish the edge mode from the bulk states. This is not possible once truncation is in place. However comparing between truncated systems one sees that the hybridization decreases as $\eta$ grows (compare $\eta=3$ and $\eta=5$). 

Fig.\ref{f5} shows the amplitude of the Bloch wave function associated to the magnon spectrum of Fig.\ref{f4} at points $q_x=0, 0.5, 0.7$ in terms of the stripe width along $\hat{y}$. For $\eta<3$ the Bloch state of the candidate edge mode is delocalized at $q_x=0$, while at $\eta\geq 3$ the Bloch function at $q_x=0$ decreases from the edge in an exponential fashion which oscillates. This is also the case for $q_x=0.5$, while at $q_x=0.7$, the edge mode is well localized. This result seems to confirm that the localization of the edge states improves as the range of dipolar interactions increases. We checked the Chern numbers of the lower and upper bulk bands of the truncated systems.   We found that both bands are topologically trivial as long as $\eta<3$. For $\eta\geq 3$ the bands Chern numbers behave as the ones shown in Table~\ref{table:1} for the system with full interactions.  
This analysis allows to conclude that  the system is able to manifest topologically protected edge modes when the dipolar interactions are truncated beyond the second nearest neighbors. However, the edge states of a truncated system are not as well localized at the edge of the sample as the edge modes of fully interacting counterparts.

Finally in order examine how  the width of stripes  affects the localization of chiral edge states,  we analyzed the magnon spectrum and chiral modes of narrow stripes with $50>N_r\geq10$ \cite{supp}. The results presented in  supplementary Fig.1 \cite{supp} indicate that in narrower stripes the edge mode does not overlap with bulk eigenvalues however it may become delocalized (this is the case in the stripe with $N_r=10$).
\\
\section{Thermomagnetic Hall transport}
\label{sec:hall}
Upon applying a temperature gradient, the magnon Hall effect MHE allows a transverse heat current mediated by magnons in two dimensions \cite{onose2010observation} which has been explained in terms of uncompensated magnon edge currents \cite{matsumoto2011theoretical,matsumoto2011rotational}. The relevant quantity characterizing the MHE is the thermal Hall conductivity which is associated to the Berry curvature of the eigenstates. The intrinsic contribution to the transverse thermal conductivity
\begin{equation}
\label{hall }
\kappa^{xy}=-\frac{k_B^2 T}{4\pi^2\hbar}\sum_i\int_{BZ}\theta(\rho_i)F^{xy}_i(\bm q)\bm dq
\end{equation}
is intimately related to the Chern numbers defined in Eq.~(\ref{eq:chern2}). The sum is over all bands $i$ in the magnon dispersion, and the integral is over the 1BZ. $\rho_i$ is the Bose distribution function and the function,
\begin{equation}
\label{theta}
\theta(x)\equiv (1+x)(\ln\frac{1+x}{x})^2-(\ln x)^2
-2Li_2(-x)
\end{equation}
where $Li_2$ is the dilogarithm, $T$ the temperature, and $k_B$ is the Boltzmann constant. The thermal Hall conductivity can be interpreted as the Berry curvature weighed by the $\theta$ function \cite{matsumoto2011rotational}. The sign of $\kappa^{xy}$  
depends on the topological phase of the bulk system. This dependence can be understood in terms of edge modes and their propagation direction \cite{mook2014edge}. The topological phases (1,-1) and (-1,1) produce one edge mode each. They differ in the slope of their dispersion. In the first case, the nontrivial edge mode propagates to the right, while in the second, it does to the left. The sign of $\kappa^{xy}$, and therefore the direction of the heat transport in a given topological phase, depends on the occupation probability of the edge magnons. When there is more than one edge mode  with different slopes in the same phase, two propagation directions are possible depending on $T$. Since $\kappa^{xy}$ is weighted by the function $\theta$, edge modes propagating in different directions may induce cancellation of the transverse thermal conductivity at high energies. If all nontrivial edge modes propagate in the same direction as happens in this case, the sign of the thermal Hall conductivity is fixed within the topological phase, and its sign does not depend on temperature. Here,  Fig.~\ref{f6} shows that phase (1,-1) at $h<h_2$ has $\kappa^{xy}>0$, while in phase (-1,1) at $h>h_2$, $\kappa^{xy}<0$. $h$ not only switch the sign of $\kappa^{xy}$ but also changes its magnitude: in phase (1,-1) reduced values of $h$ tend to increase the magnitude of the transverse thermal conductivity.  \\
\section{Effective model near band touching points.}
\label{sec:effective}
The band touching at $p_1=(q_0,h_1)$ shown in Fig.~\ref{f3}(b) can be seen as a single Dirac point around which the frequency dispersion for both bands can be approximated by a linear function 
\begin{eqnarray}
\omega^{(p_1)}_{1,2}\sim \pm\nu|q|\nonumber
\end{eqnarray}  
with $\nu$ the speed of magnons in the chain. The singular structure of the frequency dispersion near the band touching can be studied using degenerate perturbation theory \cite{shindou2013chiral}. For the magnon hamiltonian of Eq.~\ref{eq:hmag}, it takes the form 
\begin{eqnarray}
H_p=H_{1}+V_p\nonumber
\end{eqnarray} 
with $H_1=\mathcal{H}_d(p_1)$ and $V_p=H_p-H_{1}$. At the touching point, $H_{1}$ has twofold degenerate eigenstates $| d_j \rangle$ (j=1,2) with eigenfrequency $\omega_0$ $(>0)$, that satisfies $H_{1}| d_j \rangle=\omega_0\sigma_z | d_j \rangle$ \cite{supp}. On introducing the perturbation $V_p$, the degeneracy is split into two frequency levels. The eigenstate for the respective eigenfrequency is determined on the zero order of $p-p_1$ as
\begin{eqnarray}
T_p=T_1U_p+\mathcal{O}(|p-p_1|)\nonumber
\end{eqnarray}   
where the matrix $T_1$ diagonalizes $H_1$ and the unitary matrix $U_p$ diagonalizes a 2 by 2 hamiltonian $h_{p}$ formed by the twofold degenerate eigenstates,
\begin{eqnarray}
h_p=\left(\begin{array}{cc}d_1^\dagger V_p d_1 & d_1^\dagger V_p d_2 \\d_2^\dagger V_p d_1 & d_2^\dagger V_p d_2\end{array}\right)\nonumber
\end{eqnarray}  
 In Fourier space this can be written as 
\begin{eqnarray}
h_q= \left(\begin{array}{cc}f_1(q,h) & f_2(q,h) \\f_3(q,h) & f_4(q,h)\end{array}\right)\nonumber
\end{eqnarray}  
where we find that $f_1(q,h)=-f_4(q,h)$ and $f_2(q,h)=f_3(q,h)$, \cite{supp}. Expanding $f_1(q,h)$ and $f_2(q,h)$ near $p_1$ \cite{supp} yields $f_1(q,h)\sim \beta(h-h_1)$ and $f_2(q,h)\sim-i\nu (q-q_0)$. Thus
\begin{eqnarray}
h_q\sim-\imath\nu (q-q_0)\sigma_x+\beta(h-h_1)\sigma_z
\end{eqnarray}
with constants $\beta>0$ and $\nu>0$ \cite{supp}. Near the band touching point the effective hamiltonian becomes
\begin{eqnarray}
\mathcal{H}_{\rm{eff}}=\omega_0\sigma_0-\imath\nu(h) (q-q_0)\sigma_x+m\sigma_z
\end{eqnarray}  
 We identify the mass term $m=\beta(h-h_1)$ which cancels out at $h=h_1$ at the band crossing point. In the presence of the mass term, the spectrum becomes gapped
 \begin{eqnarray}
\omega_{1,2}\sim \pm\sqrt{(\nu q)^2+m^2}\nonumber
\end{eqnarray} 

\section{Conclusion}
\label{sec:conclusion}
In this work we demonstrate that dipolar zig-zag chains, 1d lattices of point dipoles with a two point basis are building blocks of 2d lattices with topological magnon bands. Stripes, built from a finite number of chains along $\hat{y}$ host topological chiral edge states along $\hat{x}$ direction. Such chiral behavior is exposed  by an effective model in 1d. Tuning $h$, allows for the explicit control of magnonic frequencies, the velocity of the chiral edges modes and the the transverse thermal conductivity. Indeed,  the spin-wave volume bands take non-zero Chern numbers $1$ and $-1$, respectively when $h>0$. These values are exchanged at $h=\frac{3}{2}$ due to two band touchings that yield two monopoles with charge +1 each. Due to the monopoles, the Berry phase acquires divergence, which triggers the exchange of the Chern numbers between the bands. This topological phase transition causes the change of sign of both the Hall conductivity and the sense of motion of edge states in the system. 
The topological character of the bulk bands survives the truncation of the dipolar interactions up to second nearest neighbor dipoles. However we find that the localization of the edge states deteriorates as the range of interactions decreases. 

The current approach for magnon manipulation leaves out the intervention of external fields and relies instead on tuning a single intrinsic geometrical parameter. The calibration of $h$ tunes the internal anisotropic magnetic fields in the lattices. Such internal fields originate from dipolar interactions between magnets and produce a spin-momentum locking, which constrains the dipole's magnetic moment orientation.  Indeed, we find that the dipolar interactions in systems that feature zig-zag point dipoles exactly maps  into symmetric plus an antisymmetric  DM spin orbit type of interaction which is consistent with previous works where the DM coupling has been introduced in the hamiltonian.

A thrilling consequence of increasing $h$ is the flattening of the lower magnon band. The effect of the band flattening on thermal conductivity has not been addressed here, but we believe it deserves special attention. With an intrinsic knob being able to tune the flatness of the spin wavebands, the present dipolar system could open up a new avenue to study correlation-driven emergent phenomena on the background of topological magnonics. 

Possible realizations of the systems presented here could be accomplished by means of molecular magnets \cite{bogani2010molecular,syzranov2014spin}, optical lattices \cite{atala2013direct} and nanomagnetic arrays made out of permalloy \cite{gartside2020current}. Especially suitable are state-of-the-art magnonic crystals fabricated out of epitaxially grown YIG films \cite{frey2020reflection}. Recent work has shown that thickness and width modulated magnonic crystals comprising longitudinally magnetized periodically structured YIG-film waveguides can manipulate magnonic gaps with advantages such as a small magnetic damping and high group velocity of the spin waves \cite{mihalceanu2018temperature}.

\section*{Acknowledgments}
The author thanks Alexander Mook and Roberto Troncoso for fruitful discussions and thanks support from Fondecyt under Grant No. 1210083.

\end{document}